\documentclass[prc,fleqn,12pt,twoside]{revtex4}

\usepackage{graphicx}

\begin{document}

\title{Direct-decay properties of Giant Resonances}
\author{M.H. Urin}
\affiliation{Kernfysisch Versneller Instituut, University of Groningen,
9747 AA Groningen, The Netherlands}
\affiliation{Moscow Engineering Physics Institute (State University),
31 Kashirskoye shosse, 115409 Moscow, Russia}

\begin{abstract}
A semi-microscopic approach based on the continuum-RPA method and a
phenomenological treatment of the spreading effect is developed and applied to
describe direct-decay properties of a few isovector giant resonances.
Capabilities of the approach to describe giant-resonance gross properties are
also checked.
\end{abstract}


\maketitle

\section{INTRODUCTION}
Along with the strength distribution and transition density the
direct-decay probabilities are also related to the main properties
of giant resonances (GRs). In particular, the partial widths (or
the partial branching ratios) for direct nucleon decay into
one-hole states of the daughter nucleus carry information about
the particle-hole structure of a given GR and also about its
coupling to the single-particle continuum and many-quasi-particle
configurations. Direct-decay properties of GRs have been studied
experimentally for a long time (see, e.g., Refs. [1-4] and
references therein). To describe systematically these properties
we developed a semi-microscopic approach, based on the
continuum-RPA (CRPA) method and a phenomenological treatment of
the spreading effect. Within this approach we described
satisfactorily the partial widths (partial branching ratios) for
direct nucleon decay of a number of GRs. They are: the isobaric
analog resonance (IAR) [5], the Gamow-Teller and isovector giant
spin-dipole resonance (GTR and IVGSDR, respectively) [6],
the isoscalar giant monopole resonance (ISGMR)[7], the isoscalar
giant dipole resonance (ISGDR) [8]. In this work we briefly describe
the approach and present recent results concerned with
semi-microscopic description of direct-decay properties of a few
isovector giant resonances [9-13].

\section{ELEMENTS OF THE APPROACH}
\subsection{Basic CRPA equations}
In our approach the CRPA equations are used in a form which is consistent with
Migdal's finite Fermi-system theory. Of particular importance is the expression
for the effective single-particle probing operator (or the effective external
field) $\widetilde V(x,\omega)$ with $x$ being the set of radial, spin-angular,
isospin variables and where $\omega$ is the excitation energy. The difference
between the effective and the bare probing operator $V(x)$ is due to
polarization effects from the quasi-particle interaction in the particle-hole
(p-h) channel. Along with the strength function $S_V(\omega)$ and transition
density $\rho(x,\omega)$ (these are related to the GR gross properties) the
effective field determines also the (direct + semidirect) nucleon-escape
amplitudes $M_{V,c}(\omega)$
\begin{equation}
M_{V,c}(\omega)\sim\int\psi_{cont}(x,\omega)\widetilde V(x,\omega)
\psi_{bound}(x)dx \label{eq1}
\end{equation}
Here, $\psi_{bound}$ and $\psi_{cont}$ are, respectively, the single-particle
bound-state (with an energy $\varepsilon_\mu$) and continuum-state (with an
energy $\varepsilon=\varepsilon_\mu+\omega$) wave functions and $c$ is a set of
decay-channel quantum numbers. Within the CRPA the unitary condition is
fulfilled: $S_V(\omega)=\sum_c\limits |M_{V,c}(\omega)|^2$. (It is supposed
that $\omega > S_N$, the latter is the nucleon separation energy).

\subsection{The spreading effect}
In the present approach the spreading effect is phenomenologically taken into
account in terms of an appropriate spreading parameter $I$. In the CRPA
description of low-energy (``sub-barrier'') GRs one can represent the reaction
amplitudes (i.e. polarizability $P_V(\omega)$ and amplitudes $M_{V,c}(\omega)$)
as a sum over non-overlaping doorway-state resonances. The spreading parameter
has the meaning of the averaged doorway-state spreading width and can be
introduced in the expansion of the reaction amplitudes by the substitution
$\omega\to\omega+(1/2)iI$ to obtain the corresponding energy-averaged
amplitudes. The parameter $I$ is adjusted to reproduce in calculations of the
energy-averaged strength function $S(\omega)=-\frac1\pi {\mathrm Im} P(\omega)$
the observed total width of a given GR. Then the partial direct-nucleon-decay
branching ratios
\[
b_c=\frac{\int |M_{V,c}(\omega)|^2d\omega}{\int S_V(\omega)d\omega}
\]
can be calculated without the use of any new parameters [6,7]. If the
doorway-state resonances are overlapping (which generally is the case for
high-energy ``over-barrier'' GRs), we take the spreading effect into account by
the above-mentioned substitution of complex energies directly in the CRPA
equations [8]. In this case the spreading parameter, taken radial- and
energy-dependent, is, actually, the twice imaginary part of the potential used
in calculations of the $\omega$-dependent single-particle quantities. Such a
potential is not directly related to the optical-model potential used in
description of nucleon-nucleus scattering (see, e.g., Ref. [14]). This
procedure allows us to calculate directly the energy-averaged characteristics
of a given GR (the strength function, energy-dependent transition density and
nucleon-escape amplitudes) while taking into account all main GR relaxation
modes.

\subsection{Input quantities and model parameters}
A realistic phenomenological isoscalar nuclear mean field (including the
spin-orbit term) and the (momentum-independent) Landau-Migdal p-h interaction
are used as input quantities for CRPA calculations. The mean field and the p-h
interaction (in the non-spin-flip channel) are related to each other by
self-consistency conditions, which are due to the basic symmetries of the model
Hamiltonian. Using these conditions we calculate self-consistently the
isovector part of the nuclear mean field (symmetry potential) via the
phenomenological Landau-Migdal parameter $f'$, the mean Coulomb field, and
determine the strengths of the p-h interaction in the isoscalar non-spin-flip
channel. The parameters of the isoscalar part of the nuclear mean field and the
parameter $f'$ are chosen such that the nucleon separation energy and
single-quasi-particle spectrum in closed-shell subsystems can be described
satisfactorily.  The strength $g'$ of the p-h interaction in the isovector
spin-flip channel is chosen such that the experimental GTR energy is reproduced
in calculations. We also take into account, in an effective way, the
(relatively small) contribution of isovector momentum-dependent forces in
formation of the isovector GRs by using a scaling transformation of the
reaction amplitudes calculated within the CRPA. The corresponding strength
parameter $k_L$ (the ``velocity'' parameter), which describes the contribution
of the momentum-dependent forces in the corresponding energy-weighted sum rule,
is taken such that description of the peak energy (and exhaustion of the total
strength) for given isovector GR is improved [9-13]. The radial dependence of
the spreading parameter is taken the same as for the isoscalar mean field,
while the energy dependence is described by a function with saturation-like
behavior (such a function is used for the imaginary part of a single-particle
potential in some versions of the optical model of the nucleon-nucleus
scattering). The parameters of this function have been found in Ref.~[8] from
the semi-microscopic description of the total width of a few isoscalar
resonances in a number of singly- and doubly-closed shell nuclei. In the
description of the isovector GRs we occasionally change the strength of
spreading parameter by a small amount to improve the description of the
experimental strength distribution [9]. It should be stressed, that after the
above-outlined choice of the model parameters we do not use any new parameters
to describe the direct-nucleon-decay properties of a given GR.

\section{APPLICATIONS}
\subsection{Direct + Semidirect (DSD) photoneutron and inverse reactions}
Since the excitation of electric GRs (EL-GRs) is an intermediate step in DSD
reactions, these reactions are closely related to the direct-decay properties
of EL-GRs. In the present approach the reaction amplitudes are proportional to
the nucleon-escape amplitudes of Eq.~(\ref{eq1}), provided that the appropriate
external field is used. For a specific nucleus we first describe the
experimental photo-absorption cross section in the energy region of the
isovector giant dipole resonance (IVGDR) to obtain the spreading parameter
strength $\alpha$ and the ``velocity'' parameter $k_1$. The values for $k_1
\simeq$ 0.1--0.2 are in agreement with the systematics of Ref.~[1], while the
extracted values for $\alpha$ are close to the value $0.125$ MeV$^{-1}$ used in
Ref.~[8]. Following this procedure the partial neutron-radiative-capture cross
sections for $^{89}$Y, $^{140}$Ce, and $^{208}$Pb target-nuclei have
satisfactorily been described without the use of new parameters [9]. Actually,
the present model can be regarded as the semi-microscopic version of the
well-known phenomenological DSD-model (see, e.g., Ref.~[1] and references
therein). In the ``single-level'' approximation one can get from
Eq.~(\ref{eq1}) the expression for the reaction amplitude in the form used
within the DSD-model. The difference is that the GR form factor is proportional
to the GR transition density and has no imaginary part. In addition the
continuum-state wave function is calculated with the use of an effective
optical-model potential.

Within the same model we evaluate the total direct-neutron-decay branching
ratio for the IVGDR in a number of singly- and doubly-closed shell nuclei [9].
The value of 21.4\%, obtained for $^{48}$Ca, is found to be only in qualitative
agreement with the corresponding experimental value of 39(5)\% [2].

As the next step, we consider the backward-to-forward asymmetry of the
differential cross sections of the above-mentioned reactions in the energy
region of the isovector giant quadrupole resonance (IVGQR) [11]. The asymmetry
is due to an interference between the E1- and E2-reaction amplitudes and,
therefore, reveals a non-monotonous energy dependence in this energy region.
For this reason, experimental studies of the asymmetry presents an indirect way
to locate the IVGDR (see Ref.~[1] and references therein). Considering as an
example the nucleus $^{208}$Pb, we satisfactorily described the experimental
data on the asymmetry (defined as the difference-to-sum ratio of the
differential cross sections taken at $55^\circ$ and $125^\circ$) using the
``velocity'' parameter value $k_2=0.1$. After this we calculate all the main
properties of the IVGQR in $^{208}$Pb without the use of free parameters. In
particular, the peak energy $\simeq 21.5$ MeV and the total width $\simeq 7$
MeV are found in agreement with the systematics of Ref.~[1].

\subsection{Overtone of the IVGDR}
Most of high-energy GRs are the next vibration modes (overtones, or
second-order GRs) relative to the corresponding low-energy GRs (main tones, or
first-order GRs). The lowest energy isoscalar second-order GR is the
well-studied ISGDR (the overtone of the zero-energy $1^-$ spurious state,
associated with the center-of-mass motion). One can expect that in the neutral
channel the lowest energy isovector second-order GR is the overtone of IVGDR
(i.e. IVGDR2). The IVGDR2 is the isovector partner of the ISGDR. Considering
$^{208}$Pb as an example, we performed a semi-microscopic analysis of the main
properties of the IVGDR2 [12]. As an external field, the second-order dipole
operator is used with a radial dependence $r(r^2-\eta)$. The parameter $\eta$
is found from the condition of ``minimum exciting'' of the main-tone resonance.
The isovector second-order dipole strength function reveals a well-formed
resonance with a peak energy $\simeq 34$ MeV and total width $\simeq 15$ MeV.
The overtone coupling to the single-particle continuum is, naturally, more
intensive than the main-tone coupling: $b_n^{tot}\simeq  5.5$\% and 48\%,
$b_p^{tot}\simeq 0$ and 19\% for the IVGDR and ISVGDR2, respectively. The
relatively large total direct-proton-decay branching ratio allows us to suppose
that the IVGDR2 can be observed in the proton-decay channel following inelastic
scattering of electrons or light ions.

\subsection{Direct proton decay of the isovector giant spin-monopole resonance}
Among the observed GRs the charge-exchange (in the $\beta^-$ channel) giant
spin-monopole resonance (IVGSMR$^{(-)}$, the overtone of the GTR) has the
highest excitation energy ($\simeq 37$ MeV in $^{208}$Bi [3]). From comparison
of the observed and calculated total width ($\simeq 14$ MeV [3] and $\simeq 11$
MeV, respectively) one can conclude that the spreading effect for this
resonance is relatively weak, while the coupling to the (single-proton)
continuum is rather strong. This observation allowed us to put forward the
assumption that the spreading effect reveals a saturation-like energy
dependence. This assumption was successfully applied in Ref.~[15], where the
partial and total direct-proton-decay branching ratios were calculated for the
IVGSMR$^{(-)}$. In particular, the calculated $b_p^{tot}$ value was found to be
close to the corresponding value deduced later from the joint analysis of the
inclusive $^{208}$Pb($^3$He,t) and coincidence $^{208}$Pb($^3$He,tp)
experiments [3]. However, the experimental distribution of the total decay
probability over the partial decay channels (which are associated with
population of single-hole states $\mu^{-1}$ in $^{207}$Pb) has been found to be
in a noticeable disagreement with predictions of Ref.~[15]. In particular, a
rather strong population of neutron deep-hole states has been unexpectedly
found [3]. Recently, we have revised the calculations of Ref.~[15] (where, in
particular, the spreading effect on the continuum-state wave function of
Eq.~(\ref{eq1}) has not been taken into account) [13]. Although the calculated
and experimental total proton branching ratios are found to be rather close
(63\% and 52(12)\% [3], respectively), the disagreement with the experimental
decay-probability distribution still remains. The reasons for this disagreement
are not clear now. Being motivated by forthcoming experimental results, we made
some predictions for decay properties of the IVGSMR$^{(-)}$ in $^{90}$Nb and
$^{120}$Sb [13]. In particular, the $b_p^{tot}$ values (72\% and 65\%,
respectively) have been obtained for these GRs.

\subsection{Direct decays of isolated $1^-$ IAR}
The isospin splitting of the IVGDR into two components takes place for nuclei
having not-too-large neutron excess $(N-Z) = 2T$ ($T$ is the value of the
ground-state isospin). The $T_> = T+1$ component of the IVGDR (i.e. IVGDR$_>$)
is the isobaric analog of the charge-exchange (in the $\beta^{(+)}$ channel)
IVGDR (i.e. of the IVGDR$^{(+)}$) and presents the specific double GR (see,
e.g., Ref.~[1]). Assuming that the isobaric analog state exhausts 100\% of the
Fermi strength, one can express the IVGDR$_>$ strength function (corresponding
to $V(x)\sim -(1/2)\tau^{(3)}$) via the IVGDR$^{(+)}$ strength function
(corresponding to $V(x)\sim\tau^{(+)}$)
\begin{equation}
S_>(\omega)=(2T+2)^{-1}S^{(+)}(\omega-\Delta_C)
\label{eq2}
\end{equation}
where $\Delta_C$ is the Coulomb displacement energy. CRPA
calculations of the $S^{(+)}$ strength function, performed in
Ref.~[10] for the $^{48}$Ca and $^{90}$Zr parent nuclei, show that in
accordance with Eq.~(\ref{eq2}) the low-energy part of the
IVGDR$_>$ strength function contains a few isolated resonances
($1^-$ IAR). Since the cross section of photo-absorption,
accompanied by excitation of the IVGDR$_>$, is proportional to
$S_>(\omega)$, one can evaluate the partial radiative width
$\Gamma_{\gamma_0}$ of each mentioned $1^-$ IAR [10].

Being motivated by the experimental results of Ref.~[16], where one of $1^-$
IAR in $^{90}$Zr (with $E_x = 16.28$ MeV) has been studied via the
(e,e'p)-reaction, the authors of Ref.~[10] attempted to evaluate the partial
protons widths of this IAR for decay into one-hole states of $^{89}$Y. They
used the expression for the partial amplitude of the ($\gamma$,p)-reaction
accompanied by excitation of IVGDR$_>$ [10]. This expression, derived using the
above-mentioned assumption regarding the properties of the isobaric analog
state, can be presented as (compare with Eq.~(\ref{eq1}), both Eqs. are given
in a rather schematic form):
\begin{eqnarray}
M_c^>(\omega) & \sim & (2T+2)^{-1}\int \psi_{cont}(x,\omega) v(x)
\times \\ \nonumber
& & \times g_n(x,x',\varepsilon')\widetilde V^{(+)}(x',\omega')\psi_{bound}(x')dx dx'.
\label{eq3}
\end{eqnarray}
Here, $v(x)$ is the symmetry potential; $g_n(x,x',\varepsilon')$
is the neutron Green's function ($\varepsilon' = \varepsilon -
\Delta_C$); $\widetilde V^{(+)}(x',\omega')$ is the effective
charge-exchange dipole field ($\omega'=\omega-\Delta_C$). The poles in the
omega-dependence of the amplitude $M_c$ correspond to the $1^-$ IAR, while the
pole residue is proportional to the product
$(\Gamma_{\gamma_0})^{1/2}(\Gamma_p)^{1/2}$. Therefore, the joint analysis of
Eqs.~\ref{eq2} and \ref{eq3} allows one to evaluate within the CRPA the partial
proton widths, $\Gamma_p$, of each $1^-$ IAR. Among these resonances in
$^{90}$Zr one ${1^-}$ IAR with the energy $\simeq 16$ MeV can be related to the
one studied in Ref.~[16]. In the Tab. 1 the partial widths calculated for this
resonance are shown in comparison with the corresponding experimental values.
To understand the role of proton pairing in the formation of the $1^-$ IAR we
take pairing into account by simply substituting in the CRPA equations the
occupation numbers by the Bogoluybov coefficients $v^2$. This can be done in an
isospin-self-consistent way [5]. The results are also shown in the Table~1 and
demonstrate a satisfactory agreement with the data of Ref.~[16].

\section{CONCLUSION}
The semi-microscopic approach presented here is relatively simple to use and at
the same time is able to give a satisfactorily description of most of several
different experimentally measurable quantities related to the direct-decay
properties of giant resonances in singly- and doubly-closed-shell nuclei. 

The current problems with developing the approach are related
to correct description of the non-resonance part of the reaction
amplitudes and taking nucleon pairing into account (in the spirit
of Ref.~[17]). \\

\par
The author would like to thank Prof. M.N.~Harakeh for interesting discussions,
many valuable remarks and warm hospitality during a long-term stay at KVI. The
author also thanks O.~Scholten and H.J.~W\"{o}rtche for their kind help in
improving the English of the manuscript. The support from the ``Nederlandse
organisatie voor wetenschappelijk onderzoek'' (NWO) is acknowledged. This work
is partially supported by the Russian Fund for Basic Researches (RFBR) under
the grant No.~06-02-016883-a.

\begin{table}
\caption{Partial widths for decay of 1$^-$ IAR in $^{90}$Zr
($E_x\simeq 16$ MeV). $\Delta_p$ is the proton energy-gap parameter.
The decay channels $p_{0-3}$ correspond to population of the 2$p_{1/2}$,
1$g_{9/2}$, 2$p_{3/2}$, 1$f_{5/2}$ one-hole states $\mu^{-1}$ in $^{89}$Y.}
\begin{tabular}{@{}llllll}
 & $\Gamma_{\gamma_0}$ (eV) & $\Gamma_{p_0}$ & $\Gamma_{p_1}$ &
$\Gamma_{p_2}$ & $\Gamma_{p_3}$ (keV) \\
\hline
$\Delta_p=0$ & 170 & 55 & -- & 25 & -- \\
$\Delta_p=1$ MeV & 110 & 55 & 1.7 & 10.5 & 0.1 \\
exp. \cite{ref16} & 108 (35) & 54 (18)\ \  & $<$ 2\ \ \  & 20 (5) \ & 7 (2) \\
\hline
\end{tabular}
\end{table}

\end{document}